\documentclass[conference]{IEEEtran}
\usepackage[utf8]{inputenc}
\usepackage{multicol}
\usepackage{graphicx}
\usepackage{todonotes}
\usepackage{xcolor}
\usepackage{comment}
\usepackage{varwidth}
\usepackage{subfig} 
\usepackage{algpseudocode}
\usepackage{enumitem}
\usepackage{zref-xr}
\usepackage{caption}
\usepackage{multirow}
\title{Observing Responses to the COVID-19 Pandemic using Worldwide Network Cameras}

\author{
    \IEEEauthorblockN{Isha Ghodgaonkar, Abhinav Goel, Fischer Bordwell, Caleb Tung,
    Sara Aghajanzadeh, \\ Noah Curran, Ryan Chen, Kaiwen Yu,
    Sneha Mahapatra,   Vishnu Banna, Gore Kao, \\ Kate Lee, Xiao Hu, Nick Eliopolous, Akhil Chinnakotla, Damini Rijhwani,
    Ashley Kim, \\ Aditya Chakraborty, Mark Daniel Ward, Yung-Hsiang Lu, George K. Thiruvathukal\IEEEauthorrefmark{2}}
    \IEEEauthorblockA{Purdue University, West Lafayette, IN, USA\\}
    \IEEEauthorblockA{\IEEEauthorrefmark{2}Loyola University Chicago, IL, USA\\}
}

\makeatletter
\let\@previousmaketitle\@maketitle
\renewcommand{\@maketitle}{

    \@previousmaketitle

    \vspace{0.1cm}

    \setcounter{figure}{0}
    
    \includegraphics[width=.50\textwidth,height = 4.5cm]{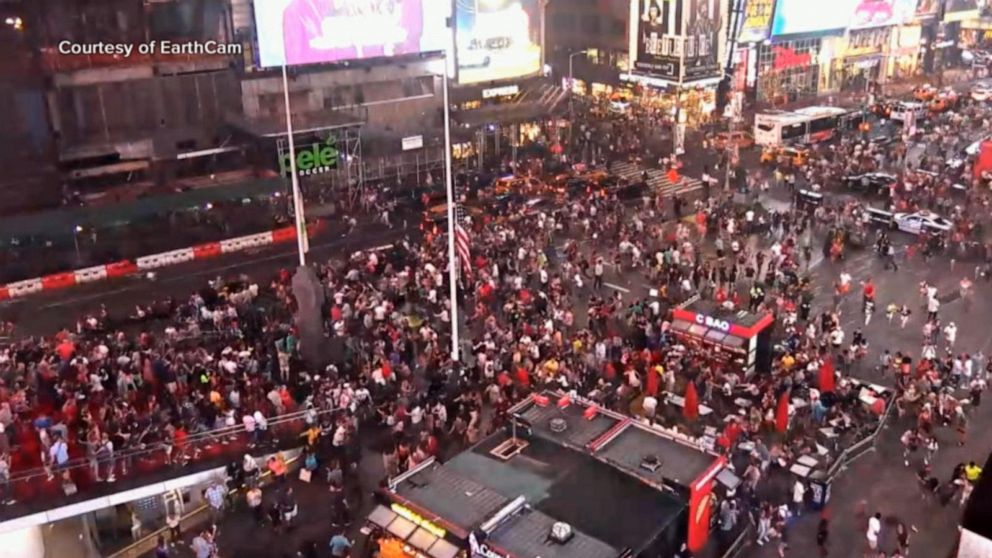}
    \includegraphics[width=.50\textwidth,height = 4.5cm]{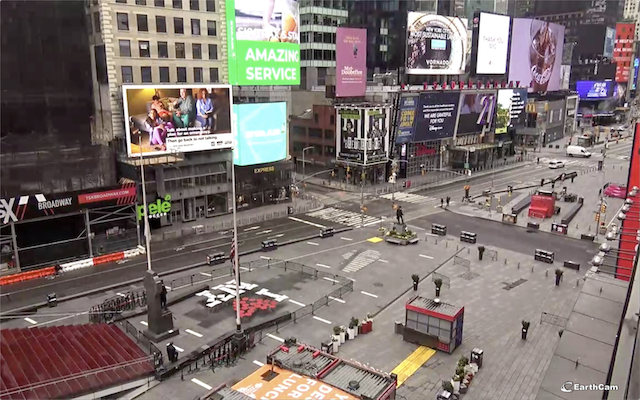}
    \captionof{figure}{Times Square, New York.  Left: 2019/08/07 (before social distancing).  Right: 2020/03/01 (during social distancing). Source: Earthcam.com}
    \label{fig:times-square-1}
    
    \vspace{0.3cm}
    
    \includegraphics[width=.50\textwidth,height = 4.5cm]{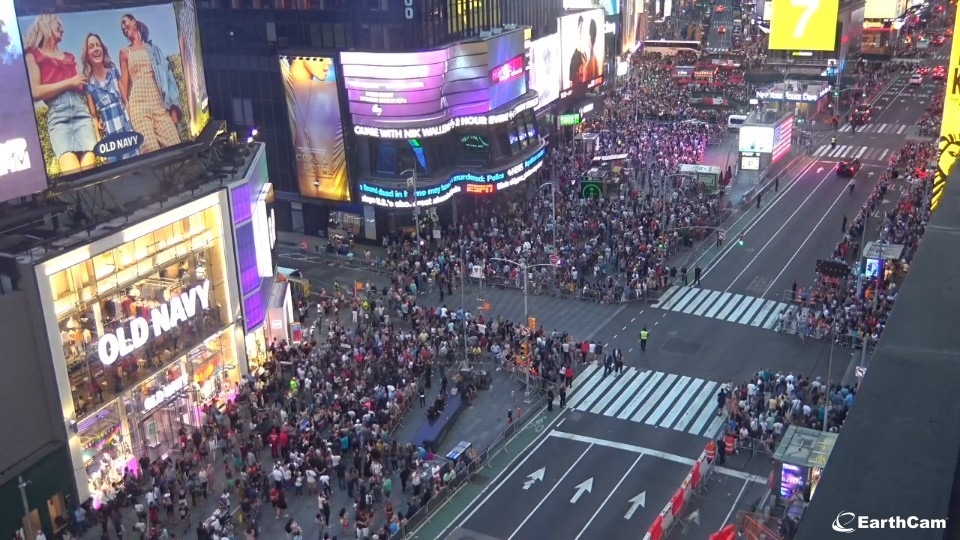}
    \includegraphics[width=.50\textwidth,height = 4.5cm]{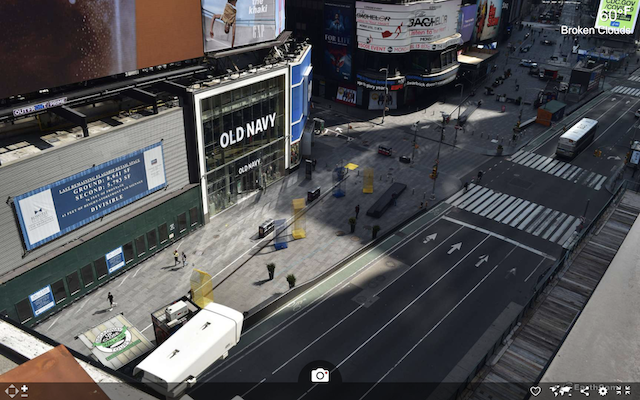}
    \captionof{figure}{Another View of the Times Square. Left: 2019/06/23. Right: 2020/03/01. Source: Earthcam.com}
    \label{fig:times-square-2}
}
\makeatother

\begin{document}

\maketitle

\begin{abstract}
COVID-19 has resulted in a worldwide pandemic, leading to ``lockdown" policies and social distancing. The pandemic has profoundly changed the world. 
Traditional methods for observing these historical events are difficult because sending reporters to areas with many infected people can put the reporters' lives in danger. New technologies
are needed for safely observing responses to these policies. This paper reports using thousands of network cameras deployed worldwide for the purpose of witnessing activities in response to the policies. The network cameras can continuously provide real-time visual data (image and video) without human efforts. Thus, network cameras can be utilized to observe activities without risking the lives of reporters. This paper describes a project that uses network cameras to observe responses to governments' policies during the COVID-19 pandemic (March to April in 2020). The project discovers over 30,000 network cameras deployed in 110 countries. A set of computer tools are created to collect visual data from network cameras continuously during the pandemic. This paper describes the methods to discover network cameras on the Internet, the methods to collect and manage data, and preliminary results of data analysis. This project can be the foundation for observing the possible ``second wave" in fall 2020. The data may be used for  post-pandemic analysis by sociologists, public health experts, and meteorologists.

\end{abstract}

\section{Introduction}
Figures \ref{fig:times-square-1} and \ref{fig:times-square-2} show the Times Square in New York City before and after the onset of the COVID-19 pandemic. Such a dramatic change is a result of social distancing, the response to this deadly pandemic. However, this change did not occur overnight. On 2019/11/17, the first known case of COVID-19 was reported in China~\cite{ma_chinas_2020}.  In January 2020, the World Health Organization (WHO) warned that the fast-spreading virus could reach other parts of the globe~\cite{nebehay_who_2020}. News media soon began to compare the coronavirus with the H1N1 influenza pandemic of 1918~\cite{grant_rebecca_2020, lovelace_jr_coronavirus_2020, barry_opinion_2020}. The 1918 influenza
infected 500 million people, about a third of the world's population at the time, and cause tens of millions of deaths. 
By March 2020, COVID-19 cases had spread around the world and had been officially labeled as a pandemic~\cite{saavedra_global_2020}. With China racing to contain the outbreak~\cite{feng_thousands_2020}, and with no vaccine available, other governments took drastic action to slow the spread of the virus. Countries near China, like South Korea~\cite{lucas_5_2020} and Taiwan~\cite{griffiths_taiwans_2020}, restricted travel to the Chinese mainland, tracked and flagged citizens via healthcare records, and enforced quarantines as deemed necessary. Italy instituted a nationwide lockdown~\cite{wallace_italy_2020}. Public travel was severely limited.  Non-essential businesses were closed, and the remaining businesses ran on restricted business hours. The United States published nationwide social distancing recommendations~\cite{harris_white_2020};  many states~\cite{saavedra_virginia_2020, bois_illinois_2020} and local governments~\cite{johnson_joint_2020} imposed stricter regulations. As a result, many businesses closed down; layoffs rocked the economy~\cite{guilford_second_2020}, and travel dwindled away as many Americans quarantined themselves in their homes.

Due to social distancing, the world has changed dramatically. With over three million confirmed cases worldwide at the end of April 2020~\cite{azner_worldwide_2020}, the COVID-19 pandemic is already a significant entry in the annals of world history. However, documenting and observing the effects of social distancing is difficult by traditional methods of journalism and observation. Dispatching photographers to a location could impose health risks. Moreover, some countries have strict travel restrictions and sending photographers
is simply not possible. It is extremely difficult for a person to take photographs at the same locations everyday for comparison during the month-long lockdown. Therefore, it is necessary to develop methods of observation that do not require physical human presence.

Thousands of network cameras have already been deployed worldwide monitoring traffic, observing national parks,
or for other purposes~\cite{see_world_through_cameras}. These cameras can be used for observing responses
of the lockdown policies. Unfortunately, network cameras are scattered over many websites on the Internet. Each website is somewhat different and uses unique methods of camera data storage and retrieval. This heterogeneity makes it difficult to find network cameras and to use them for observing responses to COVID-19.

We have developed a solution to discover network cameras and retrieve data from them~\cite{automated_discovery}. The process of automatic camera discovery is composed of a web crawler  and a module for identifying network cameras. This module distinguishes live data from unchanged images posted on websites. The team has discovered over 30,000 cameras in 110 countries across 6 continents. An image archiver is created in order to retrieve the most recent snapshot from each camera. This archiver can collect images at specified intervals. The archiver can run in parallel and capture images from multiple cameras  simultaneously. A different set of computer tools is created to handle streaming video. The project started recording data in early March at the rate of 0.5TB per week. 

This paper has the following major contributions: (1) This paper reports methods to discover network cameras on the Internet. (2) The team creates computer tools that can collect visual data (image and video) periodically 
from the thousands of network cameras. (3) This paper reports preliminary analysis showing how people respond to the policies of social distancing.

\section{Camera Discovery}
Network cameras are connected to the Internet and can provide live visual data (image and video) without human efforts. In order to observe the worldwide responses to COVID-19, thousands of network cameras are needed. Even though 
governments, universities, tourist attractions have deployed network cameras and made the data available to the public,
the data is organized in different styles and retrieval needs different protocols. 

\begin{figure}[htb!]
     \centering
     \includegraphics[width=3in]{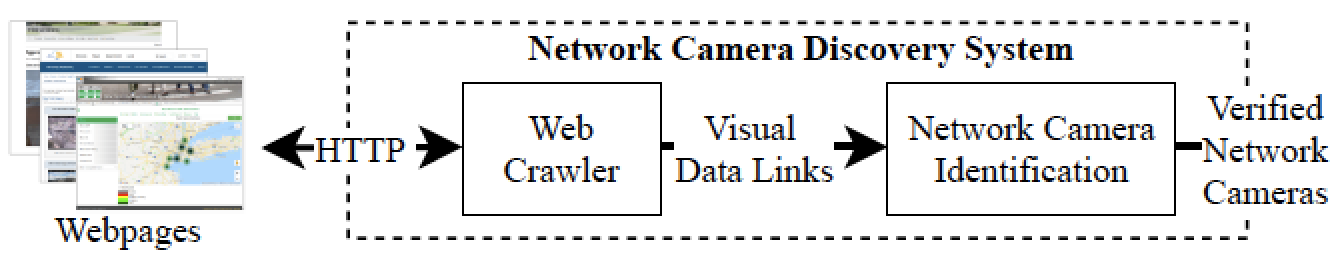}
     \caption{Automatic Network Camera Discovery System \cite{automated_discovery}. The system has 2 main parts: (1) Web Crawler module and (2) Network Camera Identification module. }
     \label{fig:Ryan_system}
 \end{figure}

This research team has created solutions for discovering network cameras automatically. The process is composed 
of two modules as shown in Figure~\ref{fig:Ryan_system}. The first module discovers websites that may contain live visual data. This is achieved by a web crawler that receives a set of seed URLs (Uniform Resource Locators) and tabulates a list of the links found from crawling the seed URLs. The module identifies the links that may contain live data and sends the those URLs to the second module. This module determines whether visual data is live or not by checking whether the data changes over time.  Different methods are needed for handling images and videos as explained below.

 \subsection{Discovery of  Image Data}

After the web crawler discovers a URL (by following links from a seed URL), the crawler 
parses the website for image formats, such as JPEG and PNG. For any visual data link found on a web page, the web crawler module downloads and parses the HTML. The pages are displayed in a web browser  environment \cite{automated_discovery} to ensure all web assets are properly loaded to avoid any potential losses of information. The crawler parses the HTML response and searches for data links common to network cameras such as image-specific visual links (e.g. baseURL /camera id.jpg) and video stream links (e.g. starting with rtmp::// and rtsp::// or ending with .mjpg). 
 
After the web crawler aggregates potential links for image data, the next step involves the network camera identification module to determine whether such data links connect to network cameras that update the data frequently. The identification distinguishes between active camera data (frequently changing) and web assets (rarely changing). The module retrieves several images from the data links at different times, and after each retrieval, compares the images to determine the change or lack thereof. The module uses three different comparison methods: (1) checksum:   compare the file checksum of the images, (2) percent difference: compare the percentage of pixels changed between images, and (3) luminance difference: compare the mean pixel luminance change between images. If the images change over time, the link is considered as connected
 to a live network camera.

 \begin{figure}
    \centering
    \includegraphics[width=3in, height=4cm]{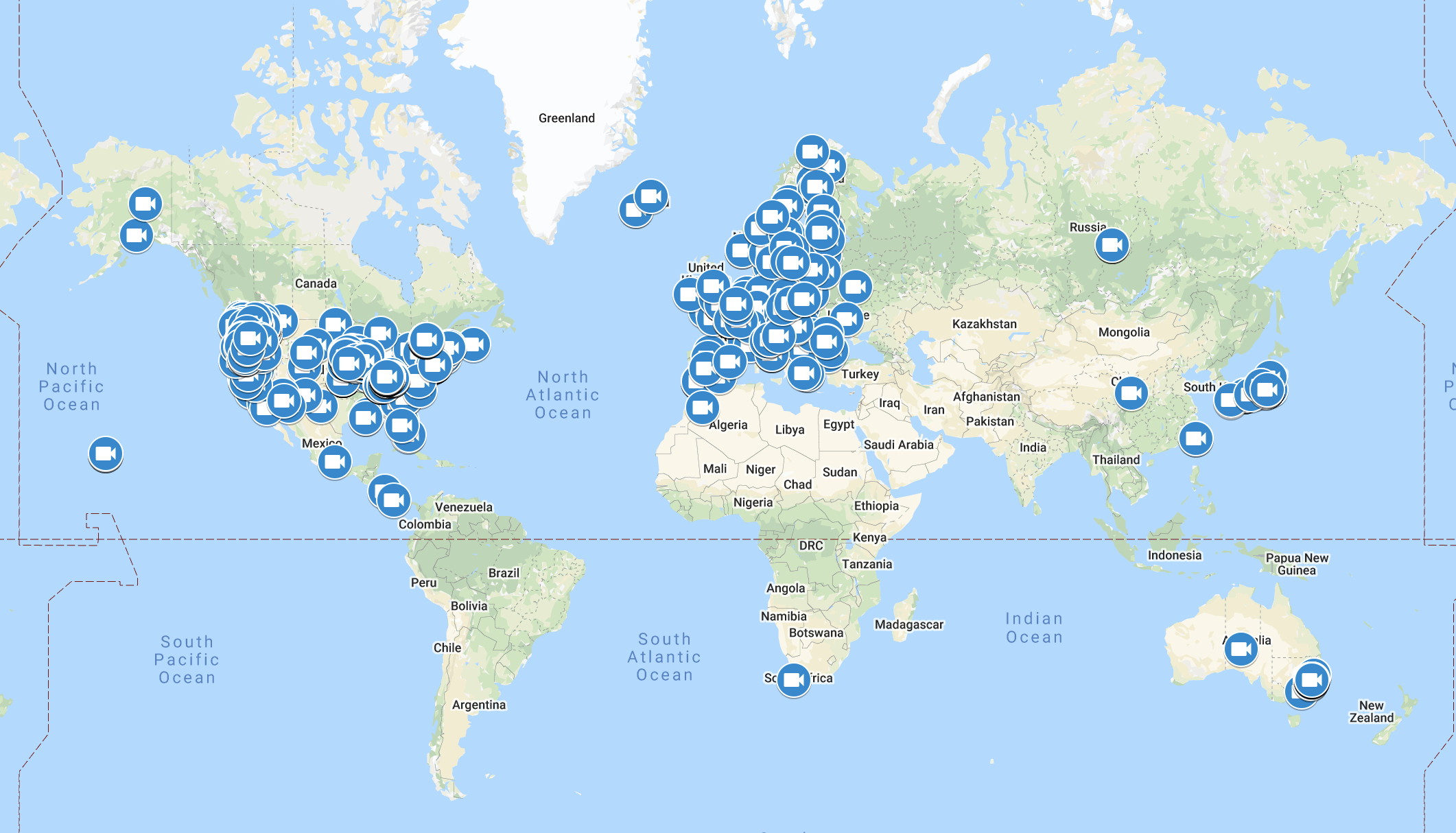}
    \caption{Map of automatically discovered live image cameras.}
    \label{fig:map1}
\end{figure}

\subsection{Discovery of Video Data}

Video data has different formats from image data. Thus, different methods are needed. This section further distinguishes two methods for discovery of video data: using web crawler or using
search engines. 

\begin{figure}
    \centering
    \includegraphics[width=3in, height = 4cm]{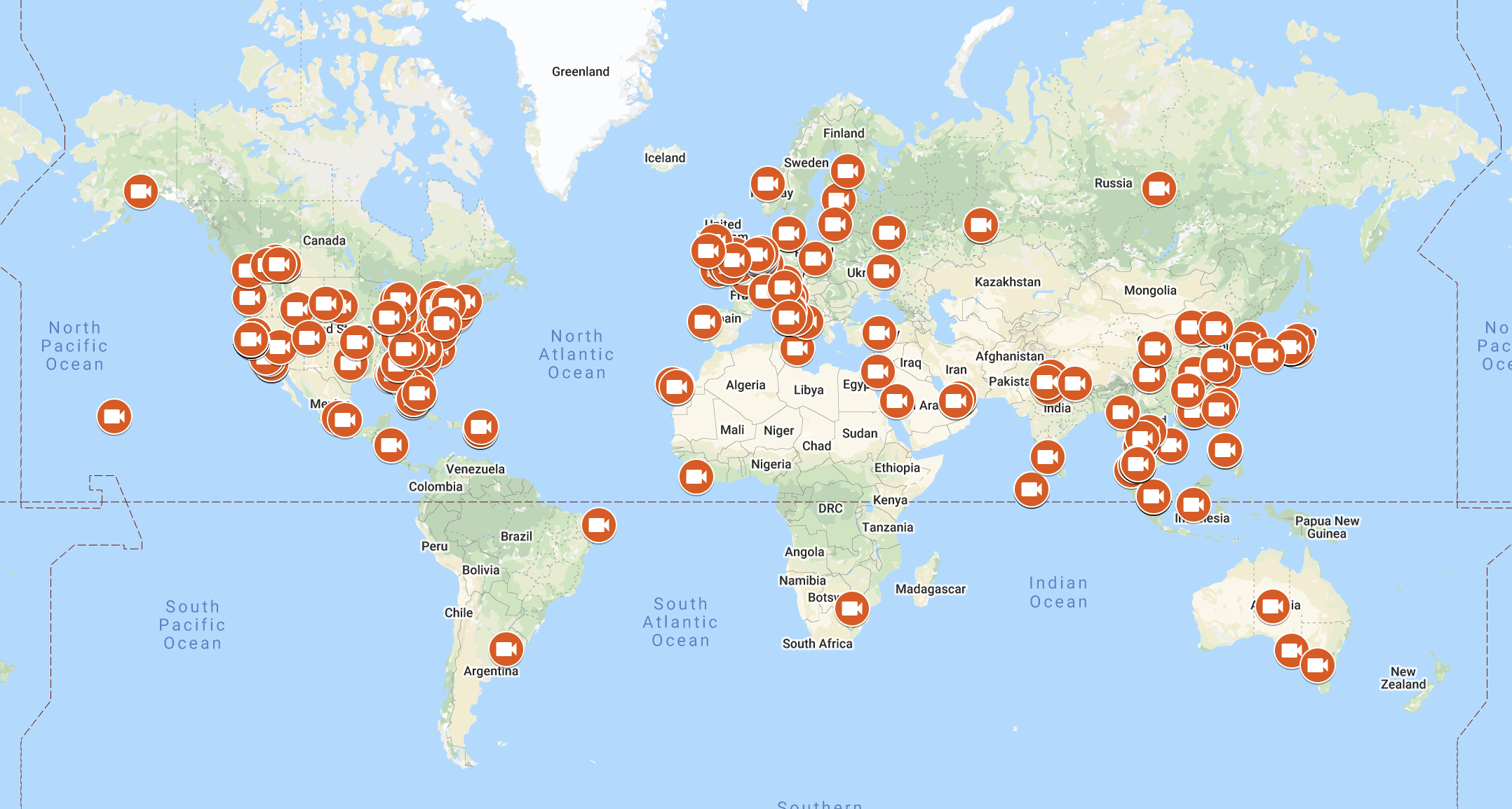}
    \caption{Map of discovered live video cameras.}
    \label{fig:map2}
\end{figure}

The first approach is based on the same method for discovering image data by replacing the identification module for recognizing live video streams. Instead of considering image formats (JPG or PNG), the crawler considers video formats: HLS, RTMP, RTSP, and MPEG. 
The identification module uses Selenium (a web browser that can simulate user interactions) to load the video stream into a Selenium Web Driver instance. Using the Selenium API, a screenshot of the video stream is taken to obtain an image. Following the same procedure as before, several screenshots are taken and compared to determine whether the stream's content changes and is likely to be a live video stream. 
Another method of finding live video streams uses an Internet search engine. More specifically, some websites, such as skylinewebcams.com, earthcam.com, and youtube.com, provide multiple live video streams discoverable by search engines, observing 
beaches, tourist attractions, or city streets. 
Approximately 400 video cameras were discovered this way. Figures~\ref{fig:map1} and~\ref{fig:map2}
show the locations of the discovered cameras.

\section{Data Collection}
\begin{figure}
    \centering
    \includegraphics[width=3in]{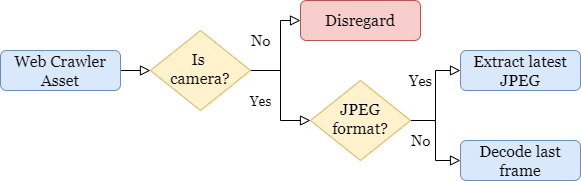}
    \caption{Flowchart of the process of extracting a frame from a camera using the Image Archiver.}
    \label{fig:flowchart}
\end{figure}

In order to compare people's responses to policy changes, it is necessary to record the data from the network cameras. This section explains how to retrieve and save the image or video data from network cameras. Figure~\ref{fig:flowchart} shows the flow of saving data.

After the network cameras are discovered, a computer program saves image data from each camera at defined time intervals (such as every 10 minutes). The program is capable of handling several data formats, including JPEG, or snapshots from MJPEG or H.264 video data. 

Video data is downloaded at regular intervals using one of several APIs for developers to access video live stream data. These APIs are built upon a video plugin system. The compatible plugin link for each video link was obtained before the relevant API methods were invoked to download data. 
    
\section{Data Management at Scale}
    
\begin{figure*}[ht]
   \centering
       \subfloat[][Canada. 7:21 AM on 2020/04/22]{\includegraphics[width=.3\textwidth, height  = 3cm]{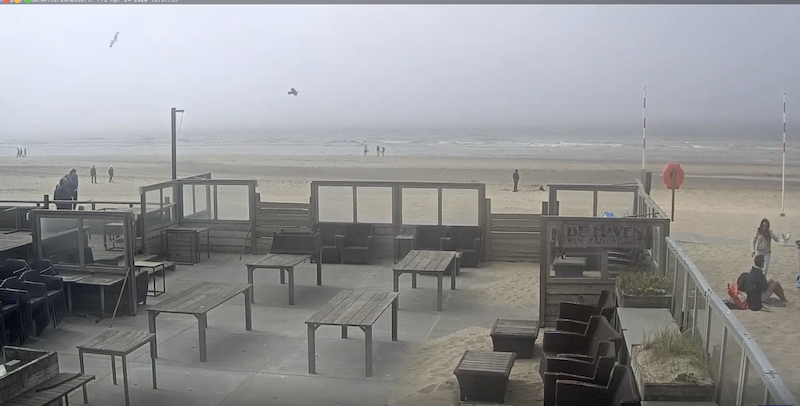}}
       \hspace{0.1mm}
       \subfloat[][Thailand. 3:20 PM on 2020/04/21]{\includegraphics[width=.3\textwidth, height  = 3cm]{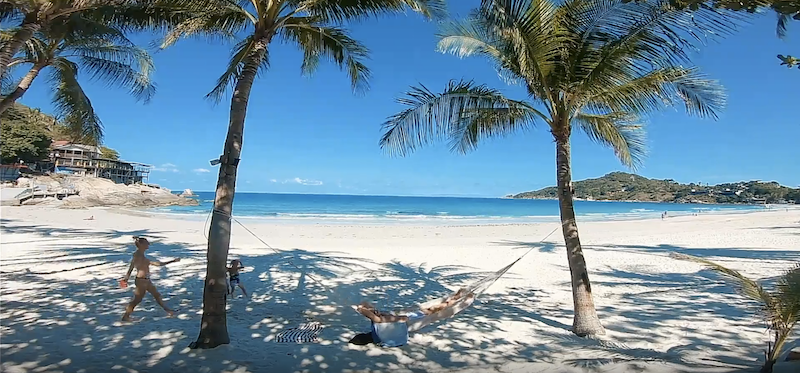}}
       \hspace{0.1mm}
       \subfloat[][Netherlands. 3:29 PM on 2020/04/24]{\includegraphics[width=.3\textwidth, height = 3cm]{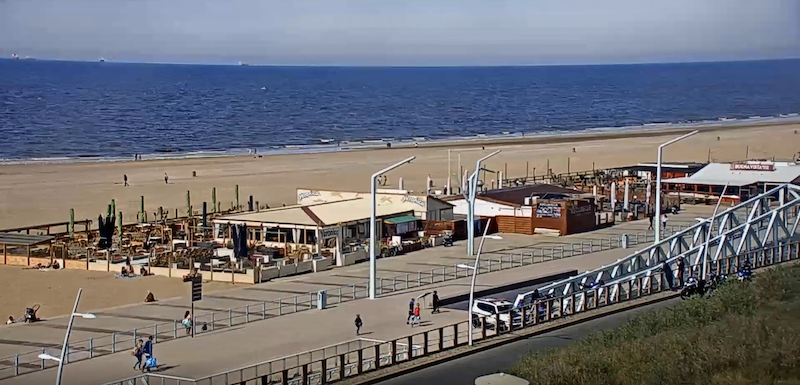}}
       \hspace{0.1mm}
   
       \subfloat[][Las Vegas, USA. 11:09 AM on 2020/04/24] {\includegraphics[width=.45\textwidth, height = 4cm ]{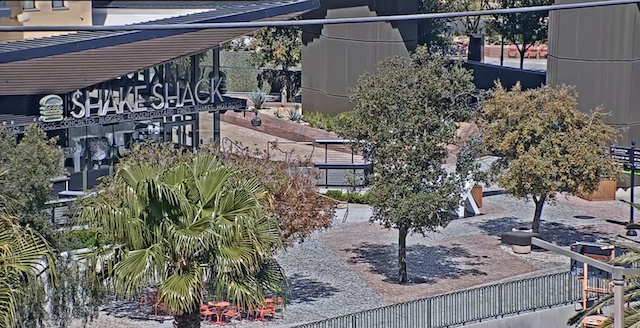}}
       \hspace{0.1mm}
       \subfloat[][Netherlands. 3:48 PM on 2020/04/24]{\includegraphics[width=.45\textwidth, height = 4cm]{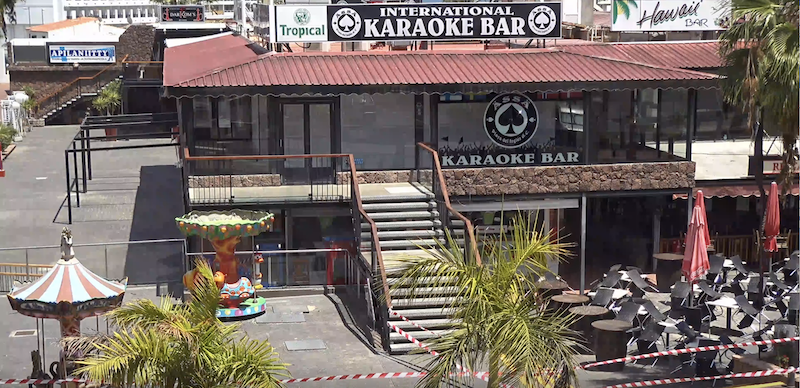}}
   \caption{Snapshots from selected network cameras around the world. All times are local time. Sources: (a) DELATECH Calgary, (b) Teleport.camera, (c) Strandweer.nu, (d) AE Signage, (e) Gran Canaria Live.}
   
   \label{fig:other}
\end{figure*}

\begin{figure*}[ht]
   \centering
   \subfloat[][2019/05/27]{\includegraphics[width=0.45\textwidth, height = 4cm]{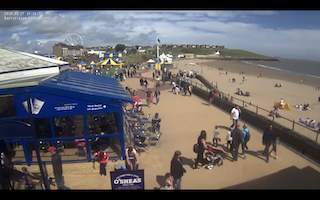}}\,
   \subfloat[][2020/04/07]{\includegraphics[width=.45\textwidth, height = 4cm]{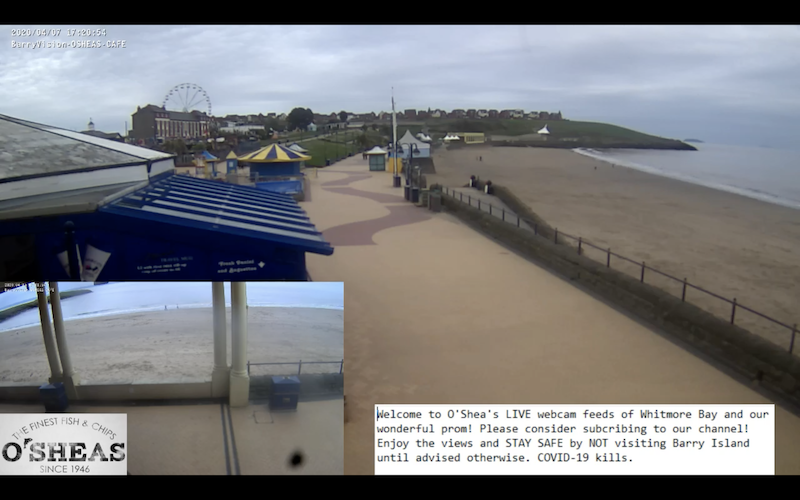}}
   \caption{Barry, Wales, United Kingdom. Source: EarthCam.com}
   \label{fig:barry}
\end{figure*}

\begin{figure*}[]
   \centering
   \subfloat[][2019/12/29]{\includegraphics[width=.45\textwidth, height = 4cm]{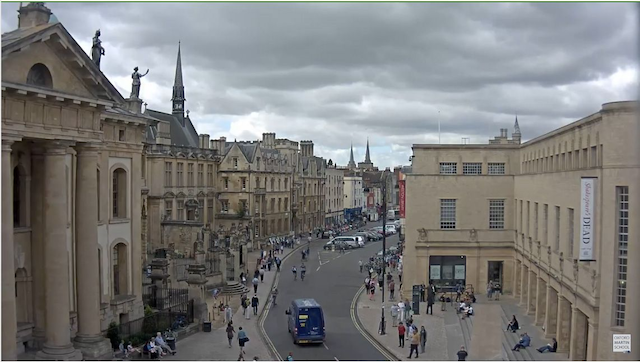}}\,
   \subfloat[][2020/05/09]{\includegraphics[width=.45\textwidth,height = 4cm]{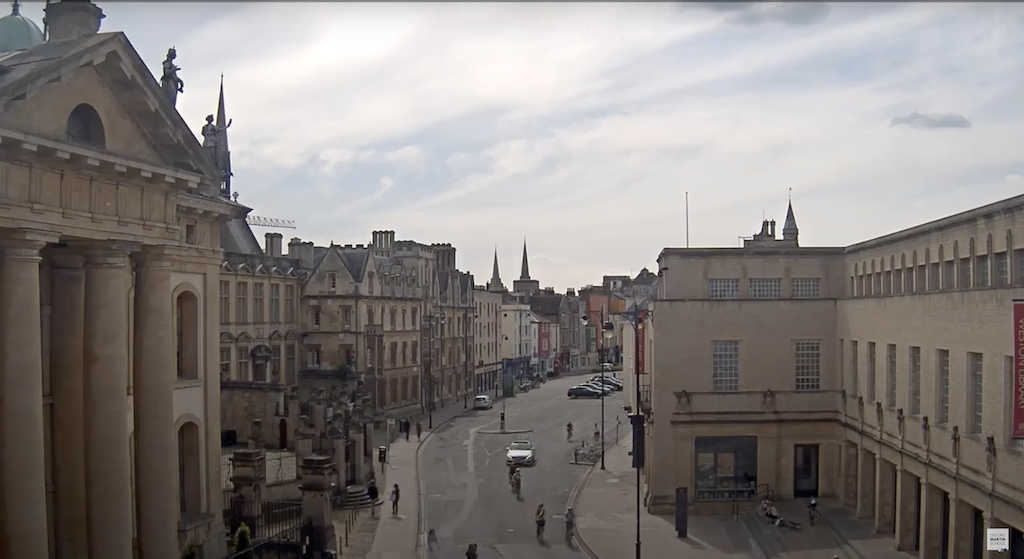}}
   \caption{Oxford, England, United Kingdom.
   Source: Oxford Martin School
   }
   \label{fig:oxford}

\end{figure*}

\begin{figure*}[ht]
   \centering
   \subfloat[][2015/03/17]{\includegraphics[width=.45\textwidth, height = 4cm]{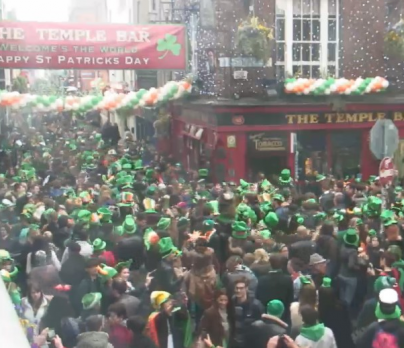}}\,
   \subfloat[][2020/03/17]{\includegraphics[width=.45\textwidth,height = 4cm]{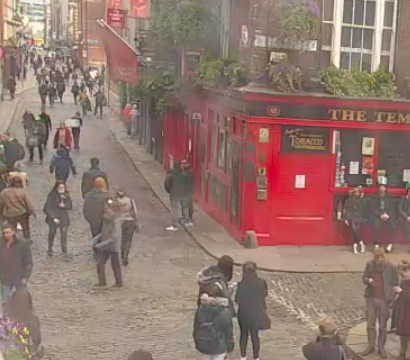}}
   \caption{
   Dublin, Ireland on Saint Patrick's Day. 
   Source: EarthCam.com
   }
   \label{fig:dublin}
\end{figure*}

\begin{figure*}[ht]
   \centering
   \subfloat[][2019]{\includegraphics[width=.45\textwidth, height = 4cm]{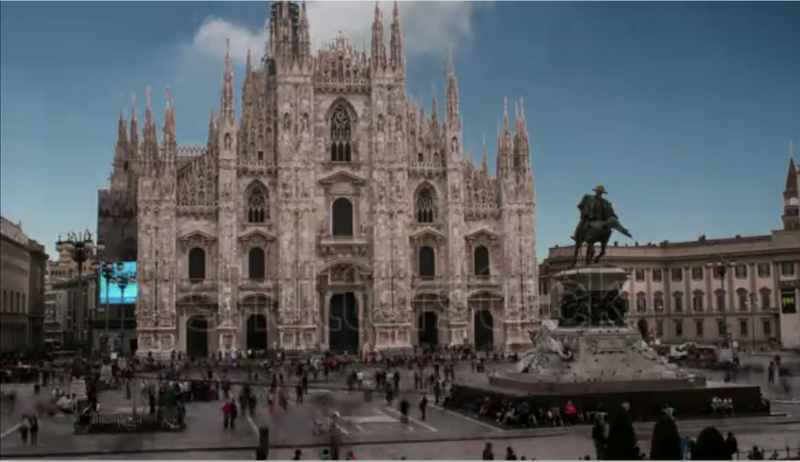}}\,
   \subfloat[][2020/04/24]{\includegraphics[width=.45\textwidth,height = 4cm]{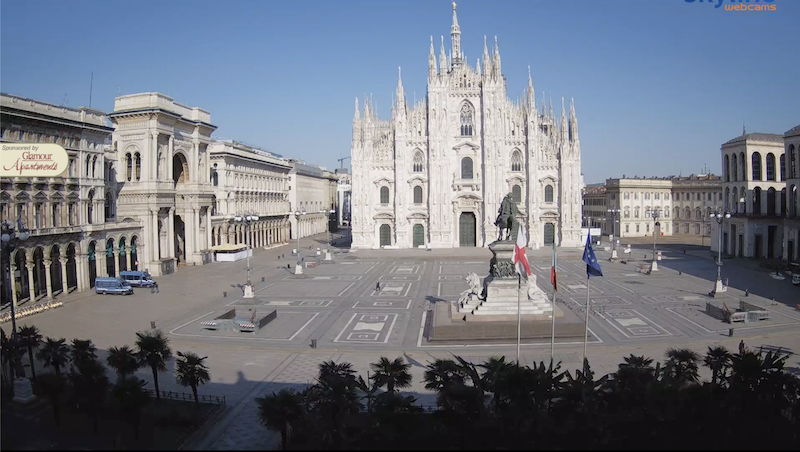}}
    \caption{Milan, Italy. Source:  shutterstock.com.
    }
   \label{fig:milan}
  
\end{figure*}

Capturing visual data from over 30,000 cameras poses two distinct challenges. The first challenge is obtaining enough computational power to collect visual data. The second challenge is running the retrieving program in parallel so that data from multiple cameras can be obtained simultaneously. This is particularly important for video because capturing streaming video requires dedicated CPU cores. Being able to overcome these two challenges is crucial to capturing visual data at scale and will become even more important as we are continuously adding cameras to our list of known cameras.

We address scalability in this work by using Cooley, a mid-sized and powerful computational cluster at Argonne Leadership Computing Facility. Through the Director's Discretionary program at Argonne, we have a renewable six-month allocation of 20,000 node-hours and 300TB of persistent storage. Cooley is primarily targeted to high-performance data visualization workloads, which also makes it ideal for capturing video capture and performing analysis.

Cooley has a total of 126 compute nodes. Each node has two 2.4 GHz Intel Haswell E5-2620 v3 processors (6 cores per CPU, 12 cores total), with 384GB RAM. This system uses FDR Infiniband interconnect. Each of the compute nodes also has an NVIDIA Tesla K80 dual-GPU card. A notable feature of our high-performance computing environment is the enormous amount of networked storage available to store data sets. At the time of writing, over 2 PB of storage is available to cluster users. Based on our current data collection rates, we will be able to collect still and video image data for more than one year.

The current focus is on using the image archiver to retrieve data from well over 30,000 cameras via a set of short jobs that run on Cooley. The image archiver itself is designed to use all available cores to fetch \emph{still} image data simultaneously. As more cameras are discovered, this will allow us to take advantage of addtional nodes without having to make any changes to how the jobs are actually submitted.

\section{Results}

    We show selected results from many popular locations around the world. We select the images shown in Figures ~\ref{fig:times-square-1}, ~\ref{fig:times-square-2}, and ~\ref{fig:barry}-\ref{fig:milan} based on two factors:
    (1) These locations are popular tourist attractions. (2) It is possible to obtain visual data before COVID-19 for comparison (this project started recording data in early March 2020).  
    Figure~\ref{fig:other} shows the diversity of the data among the discovered network cameras, including karaoke bars and beaches. 
    
    Among the data captured by more than 30,000 cameras, we choose these images as samples based on a set of criteria. All selected images observe locations where large crowds are expected because 
    they are popular tourist attractions (such as beaches) or places where gathering is expected (such as restaurants).
     Also, the large number of network cameras suggests that some may be disconnected occasionally. 
    For comparison in March and April 2020, we select the network cameras which reliably obtain data every day because they can provide insight for conducting analysis of social distancing. The criteria narrow down the number of network cameras to around 100 cameras. Then, we select cameras in different 
    geographical locations and types of location (such as beaches and restaurants).
    
Figure~\ref{fig:times-square-1}(a) and~\ref{fig:times-square-2}(a) show images of the Times Square in New York City, USA in  2019. It can be seen that the sidewalks were filled with people. Figure~\ref{fig:times-square-1}(b) and~\ref{fig:times-square-2}(b) taken in 2020 show empty streets and sidewalks because of the enforced social distancing guidelines. A similar observation can be made outside the Milan Cathedral in Milan, Italy, in figure~\ref{fig:milan}. An image from 2019 shows that the cathedral attracted large crowds of people. In 2020, due to the lockdown in Italy, there are no people visible. 

    Figure~\ref{fig:barry} shows a network camera overlooking a cafe in Barry, Wales. The beach in Barry is a popular tourist attraction in the months of April and May. Figure~\ref{fig:barry}(a) is an image taken in April 2019 and shows a crowded beach. Figure~\ref{fig:barry}(b) is taken in April 2020: all activity on the beach has been disallowed because of the enforced social distancing guidelines in Wales. A similar observation can be made in Figure~\ref{fig:oxford} in Oxford, England. Here, a camera overlooking a popular city center shows the difference in the activities between April 2019 and 2020.

    Saint Patrick's Day is a holiday in Ireland which is widely celebrated in the city of Dublin. Figure~\ref{fig:dublin}(a) shows an image from a network camera taken on Saint Patrick's Day 2015. 
    The streets were crowded and bustling with activity. On Saint Patrick's Day in 2020, social distancing measures were enforced and significantly fewer people celebrated the holiday. 
    Additional results are shown in figure~\ref{fig:other}. Figure~\ref{fig:other}(a) shows a beach in Calgary, where on the sand, there can be seen a few people at varying distances from the camera and from each other. In Figure~\ref{fig:other}(b), there appears to be a family walking past a beach goer in a hammock. In Figure~\ref{fig:other}(c), there are quite a few people walking along the beach. In Figure~\ref{fig:other}(d) and (e) there are no people, although the images show a Shake Shack, which is a popular burger joint, and a Karaoke Bar, both during the day. In each of the snapshots, there are no visible crowds. 
    These are some of the comparisons that are made possible with the use of network cameras. These examples are representative of our data set. We observe a dramatic decrease in the number of people on the streets in most countries around the world.

\section{Discussion}
     These image comparisons are made possible with the use of network cameras. We observe a dramatic decrease in the number of people on the streets in most countries around the world from Spring 2019 and earlier, to Spring 2020. We continue observing what may happen in these locations over time after social distancing policies are lifted.

    Network cameras can be useful to analyze changes  of behavior due to COVID-19, or provide valuable data for sociologists and health experts for post-pandemic analysis. For example, do people keep longer distances in crowds? Moreover, a study cross-referencing social media and public surveillance camera data can be used in emergencies~\cite{cross_reference_social}. The data set collected for this paper can be useful for researchers  for a variety of analyses.   The data may also be used to evaluate
    whether existing technologies of computer vision are ready to analyze the vast amounts of data 
    for future pandemics. In the near future, we wish to answer the following question: Is computer vision ready to analyze crowd behavior such as walking speeds, crowd density, and mask usage? Data from network cameras is significantly different from commonly used data sets due to ambient lighting; the data usually contains many objects, each with a few pixels~\cite{large_scale_obj_detect}.
    
    This research project follows the guidelines set by Purdue University for protecting privacy, approved by Purdue's Institutional Review Board Protocol 2020-460. This study analyzes only the aggregate information about crowd density and does not identify any individuals.

\section{Conclusion}

This paper presents the collection of live visual data from worldwide network cameras for monitoring responses to social distancing policies. By comparing the crowds before and during COVID-19, the data reveals how people around the world respond. This research team continuously accumulates data and plans to observe the responses when the policies are lifted.
Readers interested obtaining the data may contact the project's principal investigator, Dr. Yung-Hsiang Lu (yunglu@purdue.edu).

\section{Acknowledgements}
The snapshots shown in this paper are attributed to the following sources: Earthcam.com, YouTube.com, Skyline Webcams, O'Shea's Cafe at Barry Island, Shuttershock.com, AE Signage, Gran Canaria Live, Strandweer.nu, DELTATECH Calgary, Umbria Webcam, Teleport.camera, and Oxford Martin School.

This research uses resources of the Argonne Leadership Computing Facility, which is a DOE Office of Science User Facility supported under Contract DE-AC02-06CH11357. We thank the Argonne Leadership Computing Facility for access to the Cooley supercomputer that was used in this study.

This project is supported in part by the National Science Foundation OAC-1535108. 
Any opinions, findings, conclusions or recommendations presented in this material are those of the authors and do not necessarily reflect the views of the National Science Foundation.

\bibliographystyle{unsrt}
\bibliography{refs/caleb-zotero-refs_edited,refs/refs}

\end{document}